\documentclass[preprint,superscriptaddress]{revtex4}
\usepackage{graphicx}
\usepackage{array}
\usepackage{amsfonts}
\usepackage{amssymb,amsmath,multirow,rotate}
\usepackage{mathrsfs}
\usepackage{booktabs}
\usepackage{threeparttable}
\usepackage{multirow}
\usepackage{epsfig}
\usepackage{threeparttable}
\usepackage{chngpage}
\usepackage{latexsym}
\usepackage{dcolumn}
\usepackage{graphics,graphicx,bm,epic,eepic,float}
\usepackage{times}
\usepackage{verbatim}
\usepackage{subfigure}
\usepackage{color}
\usepackage{tabularx}
\newcolumntype{Z}{>{\centering\arraybackslash}X} 
\definecolor{red}{rgb}{1,0,0}

\definecolor{blue}{rgb}{0,0,1}

\begin{document}

\title{Synchronization of phase oscillators with
frequency-weighted coupling }

\date{\today}

\author{Can Xu}
\affiliation{College of Information Science and Engineering, Huaqiao University, Xiamen 361021, China}
\affiliation{Department of Physics and the Beijing-Hong Kong-Singapore Joint Center for Nonlinear and Complex Systems (Beijing), Beijing Normal University, Beijing 100875, China}

\author{Yuting Sun}
\affiliation{Department of Physics and the Beijing-Hong Kong-Singapore Joint Center for Nonlinear and Complex Systems (Beijing), Beijing Normal University, Beijing 100875, China}

\author{Jian Gao}
\affiliation{Department of Physics and the Beijing-Hong Kong-Singapore Joint Center for Nonlinear and Complex Systems (Beijing), Beijing Normal University, Beijing 100875, China}

\author{Tian Qiu}
\affiliation{Department of Physics, East China Normal University, Shanghai 200241, China}

\author{Zhigang Zheng}\email{zgzheng@hqu.edu.cn}
\affiliation{College of Information Science and Engineering, Huaqiao University, Xiamen 361021, China}

\author{Shuguang Guan}\email{guanshuguang@hotmail.com}
\affiliation{Department of Physics, East China Normal University, Shanghai 200241, China}

\begin{abstract}
{\bf Recently, the first-order synchronization transition has been studied in systems of coupled phase oscillators. In this paper, we propose a framework to investigate the synchronization in the frequency-weighted  Kuramoto model with all-to-all couplings. A rigorous mean-field analysis is implemented to predict the possible steady states. Furthermore, a detailed linear stability analysis proves that the incoherent state is only neutrally stable below the synchronization threshold.
Nevertheless, interestingly, the amplitude of the order parameter decays exponentially (at least for short time) in this regime, resembling the Landau damping effect in plasma physics. Moreover, the explicit expression for the critical coupling strength is determined by both the mean-field method and linear operator theory. The mechanism of bifurcation for the incoherent state near the critical point is further revealed by the amplitude expansion theory, which shows that the oscillating standing wave state could also occur in this model for certain frequency distributions. Our theoretical analysis and numerical results are consistent with each other, which can help us understand the synchronization transition in general  networks with heterogenous couplings.}
\end{abstract}

\maketitle
Synchronization in dynamical systems of coupled oscillators
is one important issue in the frontier of nonlinear dynamics and complex systems. This study provides insights for understanding the collective behaviors in many fields, such as the power grids, the flashing of fireflies, the rhythm of pacemaker cells of the heart,  and even some social phenomena~\cite{kuramoto1984chemical,strogatz2000kuramoto,pikovsky2001synchronization,arenas2008synchronization}. Theoretically, the classical Kuramoto model with its generalizations turn out to be paradigms for synchronization problem, which have inspired a wealth of works because of both their simplicity for mathematical treatment and their relevance to practice \cite{acebron2005kuramoto,dorogovtsev2008critical}.

Recently, the first-order synchronization transition in networked Kuramoto-like oscillators has attracted much attention.
For instance, it has been shown that
the positive correlation of frequency-degree in the scale-free network, or
a particular realization of frequency distribution of oscillators in an all-to-all network,  or certain special couplings among oscillators, etc, would cause a discontinuous phase transition to synchronization \cite{pazo2005thermodynamic,gomez2011explosive,leyva2012explosive,li2013reexamination,peron2012explosive,ji2013cluster,leyva2013explosive,peron2012determination,coutinho2013kuramoto,zou2014basin,skardal2014disorder,bi2014explosive,chen2015self,ji2014analysis,zhang2015explosive,ji2014low,zhang2014explosive,pinto2015explosive,yoon2015critical}. In particular, our recent work~\cite{xu2015explosive} analytically investigated the mechanism of the first-order phase transition on star network. We revealed that the structural relationship between the incoherent state and the synchronous state leads to different routes to the transition of synchronization. Furthermore, it has been shown that the generalized Kuramoto model with frequency-weighted coupling can  generate first-order synchronization transition in general networks \cite{zhang2013explosive,leyval2013explosivesy}. In Ref. \cite{hu2014exact}, the critical coupling strength for both forward and backward transitions, as well as the stability of the two-cluster coherent state, have been further  determined  analytically for typical frequency distributions.

In this paper, we present a complete framework to investigate the synchronization in the frequency-weighted Kuramoto model with all-to-all couplings. It includes three separate analyses from different angles, which together presents a global picture for our understanding of the synchronization in the model.
First,
a rigorous mean-field analysis is implemented where the possible steady states of the model are predicted, such as the incoherent state, the two-cluster synchronous state, and the traveling wave state. It is shown that in this model the mean-field frequency is not necessarily equal to 0. Instead, the non-vanishing  mean-field frequency plays a crucial role in determining the critical coupling strength.
Second,
a detailed linear stability analysis of the incoherent state is performed.
Also, the exact expression for the critical coupling strength is obtained, which is consistent with the results of the mean-field analysis, and keeps the same form for general heterogenous couplings
\cite{paisson2007synchronization,paissan2008sychronization}. Furthermore, it has been proved that the linearized operator has
no discrete spectrum when the coupling strength is below a threshold. This implies that in this model
the incoherent state is only neutrally stable below the synchronization threshold. Interestingly, numerical simulations demonstrate that
in this neutrally stable regime predicted by the linear theory, the perturbed order parameter decays to zero  and its decaying envelope follows exponential form for short time.
Finally,
a nonlinear center-manifold reduction to the model is conducted, which reveals the local bifurcation mechanism of the incoherent state near the critical point.
As expected, the non-stationary standing wave state could also exist in this model with certain frequency distributions.
Extensive numerical simulations have been carried out to verify our theoretical analyses.
In the following, we report our main results, both theoretically and numerically.

\bigskip
\noindent{\large\bf Results}\\
\noindent{\bf The mean-field theory.}
We start by considering the frequency-weighted Kuramoto model \cite{zhang2013explosive,hu2014exact}, in which the dynamics of phase oscillators are governed by the following equations
\begin{equation}\label{equ:re1}
\dot\theta_{i}=\omega_{i}+\dfrac{K|\omega_{i}|}{N}\sum_{j=1}^{N}\sin(\theta_{j}-\theta_{i}),\qquad i=1,2,\cdots,N,
\end{equation}
where $K$ denotes the coupling strength, and $\omega_{i}$ is the natural frequency of the $i$th oscillator. Without loss of generality, the natural frequencies $\{\omega_{i}\}$ satisfy certain density function $g(\omega)$ that is assumed to be symmetric and centered at $0$ throughout the paper. The most important characteristic of this generalized Kuramoto model is to introduce a frequency weight to the coupling, which leads to heterogeneous interactions in networks.
Eq.~(\ref{equ:re1}) exhibits a transition to synchronization as the coupling strength $K$ increases above a critical threshold $K_{c}$. Typically, the collective behavior in Eq.~(\ref{equ:re1}) can be characterized by the
order parameter
\begin{equation}\label{equ:re2}
\eta(t)= R(t)e^{i\Theta(t)}=\dfrac{1}{N}\sum_{j=1}^{N}e^{i\theta_{j}(t)}.
\end{equation}
Here, $\eta$ is the average complex amplitude of all oscillators on the unit circle.
$R$ is the magnitude of complex amplitude
characterizing the level of synchronization,
and $\Theta$ is the phase of the mean-field
corresponding to the peak of the distribution of phases.
When $K$ is small enough, $R(t)\approx 0$
characterizing the incoherent state in which
the phases of oscillators are almost randomly distributed.
As $K$ increases, usually a cluster of phase-locked oscillators appear, characterized by an order parameter $0<R(t)<1$. Then the system is in the  synchronous (coherent) state where the phase-locked oscillators coexist with the phase-drifting ones.

One central issue in the study of synchronization is
to identify all the possible  asymptotic
coherent states of the system as the coupling strength $K$
increases. To this end, the self-consistence method turns out to be effective.
In the following, we conduct theoretical analysis to Eq.~(\ref{equ:re1}) based on this method.

Substituting Eq.~(\ref{equ:re2}) into Eq.~(\ref{equ:re1}), we obtain the dynamical equation of the mean-field form
\begin{equation}\label{equ:re25}
\dot{\theta_{i}}=\omega_{i}+K|\omega_{i}|R\sin(\Theta-\theta_{i}).
\end{equation}
We assume that the mean-field phase $\Theta$ rotates uniformly with frequency $\Omega$, i.e., $\Theta(t)=\Omega\,t+\Theta (0)$. Without loss of generality, $\Theta(0)=0$ after an appropriate time shift. In the rotating frame with frequency $\Omega$, we introduce the phase difference
\begin{equation}\label{equ:re26}
\varphi_{i}=\theta_{i}-\Theta,
\end{equation}
and Eq. (\ref{equ:re25}) can be transformed into
\begin{equation}\label{equ:re27}
\dot{\varphi}_{i}=\omega_{i}-\Omega-KR|\omega_{i}|\sin\varphi_{i}
\end{equation}
in the rotating frame. It should be pointed out that $\Omega=0$ when the coupling form is uniform and $g(\omega)$ is even and unimodal. However, for more general cases, any asymmetry of the system, such as asymmetric frequency distribution, or asymmetric coupling function (phase lag or time delay), or asymmetric coupling strength (heterogeneous coupling or time varying coupling), etc, will cause
$\Omega \ne 0$. Therefore, there is no guarantee that $\Omega$ in Eq.~(\ref{equ:re27}) is necessarily equal to $0$.
Actually, there are several macroscopic characteristic frequencies for all oscillators, for example,
the average frequency of oscillators $\bar{\omega}=\int_{-\infty}^{\infty}\omega g(\omega)d\omega$, the mean-field frequency $\Omega$, and the mean-ensemble frequency (or effective frequency) $f_{ens}=\dfrac{1}{N}\sum_{i=1}^{N}\dot{\theta}_{i}$ \cite{basnarkov2008kuramoto,petkoski2013mean}.

Since we are interested in the steady coherent states of the system, Eq.~(\ref{equ:re27}) should be discussed in two situations corresponding to the phase-locked oscillators and the drifting ones, respectively.
On the one hand, when $|\omega_{i}-\Omega|\leq KR|\omega_{i}|$,
Eq. (\ref{equ:re27}) has solution of fixed point, i.e.,
$\dot{\varphi}_{i}=0$, which leads to
\begin{equation}\label{equ:re28}
\sin\varphi_{i}=\dfrac{\omega_{i}-\Omega}{KR|\omega_{i}|},
\end{equation}
corresponding to the phase-locked oscillators entrained by the mean-field.  On the other hand, for those drifting oscillators, $|\omega_{i}-\Omega|>KR|\omega_{i}|$.
Taking into account both the phase-locked and the drifting oscillators, the order parameter in Eq.~(\ref{equ:re2}) can be rewritten as
\begin{equation}\label{equ:re29}
\begin{split}
R=&\dfrac{1}{N}\sum_{j=1}^{N}e^{i(\theta_{j}-\Theta)}=\dfrac{1}{N}\sum_{j=1}^{N}e^{i\varphi_{j}}\\ =&\dfrac{1}{N}\sum_{j=1}^{N}e^{i\varphi_{j}}H\left(1-\left|\dfrac{\omega_{j}-\Omega}{KR|\omega_{j}|}\right|\right)+
\dfrac{1}{N}\sum_{j=1}^{N}e^{i\varphi_{j}}H\left(\left|\dfrac{\omega_{j}-\Omega}{KR|\omega_{j}|}\right|-1\right),
\end{split}
\end{equation}
where$H(x)$ is the Heaviside function. In the thermodynamical limit $N\rightarrow \infty$, the summation over the frequency should be replaced by the integration. As a result, the contribution of the phase-looked oscillators to the order parameter $R$ reads
\begin{equation}\label{equ:re30}
\int_{-\infty}^{\infty}\sqrt{1-\left(\frac{\omega-\Omega}{KR\omega}\right)^{2}}g(\omega)H\left(1-\left|\frac{\omega-\Omega}{KR\omega}\right|\right)d\omega
+i\int_{-\infty}^{\infty}\frac{\omega-\Omega}{KR|\omega|}g(\omega)H\left(1-\left|\frac{\omega-\Omega}{KR\omega}\right|\right)d\omega.
\end{equation}

In contrast to the phase-locked oscillators, the drifting oscillators could not be entrained by the mean-field.
In the thermodynamic limit $N\rightarrow \infty$, Eq.~(\ref{equ:re1}) is equivalent to the following continuity equation as a consequence of the conservation of the number of oscillators, i.e.,
\begin{equation}\label{equ:re3}
\dfrac{\partial \rho}{\partial t}+\dfrac{\partial}{\partial \theta}(\rho\cdot\upsilon)=0.
\end{equation}
Here $\rho(\theta, t, \omega)d\theta$ gives the fraction of oscillators of natural frequency $\omega$ which lie between $\theta$ and $(\theta+d\theta)$ at time $t$ with the appropriate normalization condition
\begin{equation}\label{equ:re4}
\int_{0}^{2\pi}\rho(\theta, t, \omega)d\theta=1,
\end{equation}
and $2\pi$ period in $\theta$.
Then the stationary distribution of the drifting oscillators in the rotating frame could be obtained explicitly as ($\partial \rho/\partial t=0$).
\begin{equation}\label{equ:re31}
\rho(\varphi,\omega)=\dfrac{C}{|\omega-\Omega-KR|\omega|\sin\varphi|},
\end{equation}
where
\begin{equation}\label{equ:re32}
C=\dfrac{\sqrt{(\omega-\Omega)^{2}-(KR\omega)^{2}}}{2\pi}
\end{equation}
is a normalization constant.
It is easy to obtain that for drifting oscillators
\begin{equation}\label{equ:re33}
\langle\cos\varphi\rangle=\int_{0}^{2\pi}\rho(\varphi,\omega)\cos\varphi d\varphi=0,
\end{equation}
and
\begin{equation}\label{equ:re34}
\langle\sin\varphi\rangle=\int_{0}^{2\pi}\rho(\varphi,\omega)\sin\varphi d\varphi=\dfrac{\omega-\Omega}{KR|\omega|}\left(1-\sqrt{1-\left(\frac{KR\omega}{\omega-\Omega}\right)^{2}}\right).
\end{equation}
Eq.~(\ref{equ:re33}) shows that the drifting oscillators have no contributions to the real part of $R$. However, their contributions to the imaginary part of $R$ should not be neglected. Substituting Eq.~(\ref{equ:re30})-  Eq.~(\ref{equ:re34}) into Eq.~(\ref{equ:re29}), the closed form of self-consistence equations take the following form. For the real part of $R$,
\begin{equation}\label{equ:re35}
R=\int_{-\infty}^{\infty}g(\omega)\sqrt{1-\left(\frac{\omega-\Omega}{KR\omega}\right)^{2}}H\left(1-\left|\frac{\omega-\Omega}{KR\omega}\right|\right)d\omega,
\end{equation}
and for the imaginary part of it
\begin{equation}\label{equ:re36}
\begin{split}
0=&\int_{-\infty}^{\infty}g(\omega)\frac{\omega-\Omega}{KR|\omega|}H\left(1-\left|\frac{\omega-\Omega}{KR\omega}\right|\right)d\omega+\\
&\int_{-\infty}^{\infty}g(\omega)\frac{\omega-\Omega}{KR|\omega|}\left(1-\sqrt{1-\left(\frac{\omega-\Omega}{KR\omega}\right)^{2}}\right)H\left(\left|\frac{\omega-\Omega}{KR\omega}\right|-1\right)d\omega.
\end{split}
\end{equation}
Eq.~(\ref{equ:re36}) is called as the phase balance equation~\cite{basnarkov2008kuramoto,petkoski2013mean}.  Eq.~(\ref{equ:re35}) and Eq.~(\ref{equ:re36}) together provide a closed equation for the dependence of the magnitude $R$ and the frequency $\Omega$ of the mean field on $K$.

We notice that $\Omega=0$ is always a trivial solution of Eq.~(\ref{equ:re36}), but it may not be the only solution. There may be more than one value for $\Omega$ that satisfies the phase balance equation. Considering $g(\omega)=g(-\omega)$, a pair of $\Omega$ with opposite sign might emerge. Define $\alpha=KR\geq 0$ and $x=(\omega-\Omega) / \omega$, $\Omega\neq 0$, Eq.~(\ref{equ:re35}) can be expressed as
\begin{equation}\label{equ:re37}
\dfrac{\alpha}{K}=\int_{-\infty}^{\infty}g\left(\frac{\Omega}{1-x}\right)\sqrt{1-\left(\frac{x}{\alpha}\right)^{2}}\frac{|\Omega|}{(1-x)^{2}}H(\alpha-|x|)dx.
\end{equation}
For the case of $\alpha>1$, to avoid divergency of Eq.~(\ref{equ:re37}), the only choice is $\Omega=0$, and Eq.~(\ref{equ:re35}) is reduced as
\begin{equation}\label{equ:re38}
R=\sqrt{1-\left(\frac{1}{KR}\right)^{2}},\quad \alpha=KR>1,
\end{equation}
which exactly corresponds to the two-cluster synchronous state in Ref. \cite{hu2014exact}. For the case of $\alpha<1$, the solution of Eq.~(\ref{equ:re36})- Eq.~(\ref{equ:re37}) can be solved numerically, which corresponds to the traveling wave state. In such a state,
the mean-field amplitude $R$ keeps stationary, whereas the mean-field frequency $\Omega$ differs from the mean of the natural frequencies. In particular, in the limit case $\alpha\rightarrow 0^{+}$, the critical coupling $K_{c}$ for the onset of synchronization reads
\begin{equation}\label{equ:re39}
K_{c}=\dfrac{2}{\pi|\Omega_{c}|g(\Omega_{c})},
\end{equation}
where $\Omega_{c}$ is the critical mean-field frequency.
Thus, following the analysis of Eq.~(\ref{equ:re39}), we can conclude that $\Omega_{c}=0$ means $K_{c}\rightarrow \infty$, which is not supported by numerical simulation.
By  Taylor expansion of Eq.~(\ref{equ:re36}), we find that $\Omega_{c}$ satisfies the following balance equation
\begin{equation}\label{equ:re40}
P\cdot\int_{-\infty}^{\infty}\dfrac{g(\omega)|\omega|}{\omega-\Omega_{c}}d\omega=0,
\end{equation}
where the symbol $P$ means the principal-value integration within the real line.
As an example, Table \ref{table:1} shows the balance equation~(\ref{equ:re40}), the critical mean-field frequency $\Omega_{c}$, and the critical coupling strength $K_{c}$ with respect to different frequency distributions. All these analytical results were supported by the previous numerical simulations~\cite{hu2014exact}.

\bigskip
\noindent{\bf The linear stability analysis.} The analysis of the mean-field theory above reveals four macroscopic steady states, including  the incoherent state ($R=0$), the traveling wave state ( $\Omega\neq 0$, $0<\alpha<1$ ), and the two-cluster synchronous states ($\Omega=0$, $\alpha>1$ ), respectively. However, a thorough stability analysis to every possible solution has not been performed due to the limitation of the mean-field method. In the following, we conduct a detailed linear stability analysis to the incoherent state because
its instability usually signals the onset of synchronization.
In particular, we will show that the critical coupling strength for synchronization can be alternatively
obtained via the linear operator theory.

The continuum limit of the order parameter, i.e., Eq.~(\ref{equ:re2}), is rewritten as
\begin{equation}\label{equ:re5}
\eta(t)=\int_{-\infty}^{\infty}d\omega\int_{0}^{2\pi} d\theta e^{i\theta}\rho(\theta,\omega,t)g(\omega),
\end{equation}
and the velocity is given by
\begin{equation}\label{equ:re6}
\upsilon=\omega+\dfrac{K|\omega|}{2i}(\eta(t)e^{-i\theta}-\eta(t)^{*}e^{i\theta}),
\end{equation}
where ``$*$" denotes the complex conjugate of $\eta(t)$. Let
\begin{equation}\label{equ:re7}
Z_{n}(t,\omega)=\int_{0}^{2\pi} e^{ni\theta}\rho(\theta,t,\omega)d\theta,
\end{equation}
be the $n$th Fourier coefficient of $\rho(\theta, t, \omega)$, then $Z_{0}(t,\omega)=1$ and $Z_{n}$ satisfies the following differential equations
\begin{equation}\label{equ:re8}
\dfrac{d Z_{n}}{dt}=n i\omega Z_{n}+\dfrac{n K|\omega|}{2}(\eta(t)Z_{n-1}-\eta(t)^{*}Z_{n+1}),\qquad  n=1,2,\cdots.
\end{equation}
From Eq. (\ref{equ:re5}), it is easy to verify that the order parameter $\eta(t)$ is the integral of $Z_{1}(t,\omega)$ with the frequency density function $g(\omega)$, and
the higher Fourier harmonics have no contribution to the order parameter. Since the incoherent state corresponds to the trivial solution $Z_{n}\equiv 0$ for $n=1,2,\cdots$, to study its stability we can consider the evolution of a perturbation away from the incoherent state. In this spirit, Eq.~(\ref{equ:re8}) can be linearized around the origin as
\begin{equation}\label{equ:re9}
\dfrac{dZ_{1}}{dt}=\left(i\omega+\dfrac{K}{2}\mathcal{P}\right)Z_{1}
=\mathcal{T} Z_{1},
\end{equation}
and
\begin{equation}\label{equ:re10}
\dfrac{dZ_{n}}{dt}=n\;i\omega Z_{n},\qquad n=2,3,\cdots,
\end{equation}
Here $\mathcal{P}$ is the operator defined as
\begin{equation}\label{equ:re11}
\mathcal{P}q(\omega)=|\omega|\int_{-\infty}^{\infty}q(\omega)g(\omega)d\omega.
\end{equation}
where $q(\omega)$ is a function in the weighted Lebesgue space,
and $\mathcal{T}$ is defined as
\begin{equation}\label{equ:re12}
\mathcal{T}=i\omega+\dfrac{K}{2}\mathcal{P}.
\end{equation}
From Eq.~(\ref{equ:re10}), it is obvious that the higher Fourier harmonics are neutrally stable to perturbation. Hence, the key is to study the spectrum of Eq.~(\ref{equ:re9}).
Following Ref. \cite{strogatz1991stability}, Eq.~(\ref{equ:re9}) has continuous spectrum on the whole imaginary axis.
For the discrete spectrum, we assume that the perturbation of the first Fourier coefficient has the form $Z_{1}(t,\omega)\propto e^{\lambda t}$. Then the self-consistent eigenvalue equation Eq.~(\ref{equ:re9}) for the operator $\mathcal{T}$ takes the form
\begin{equation}\label{equ:re13}
\int_{-\infty}^{\infty}\dfrac{|\omega|}{\lambda-i\omega}g(\omega)d\omega=\dfrac{2}{K},\qquad \lambda\in C\setminus i\omega,
\end{equation}
where $\lambda$ is the complex eigenvalue of $\mathcal{T}$ except for those points $i\omega$. Notice that Eq.~(\ref{equ:re13}) relates implicitly the coupling strength $K$ with the eigenvalue $\lambda$. Since the  real part of the eigenvalue $\lambda$ determines the stability of the incoherent state,
we rewrite Eq.~(\ref{equ:re13}) into two equations
by letting $\lambda=x+iy$, i.e.,
\begin{equation}\label{equ:re14}
\int_{-\infty}^{\infty}\dfrac{x}{x^{2}+(\omega-y)^{2}}|\omega|g(\omega)d\omega=\dfrac{2}{K},
\end{equation}
and
\begin{equation}\label{equ:re15}
\int_{-\infty}^{\infty}\dfrac{\omega-y}{x^{2}+(\omega-y)^{2}}|\omega|g(\omega)d\omega=0.
\end{equation}
From Eq.~(\ref{equ:re14}), we see that $x$, i.e., the real part of $\lambda$,  can never be negative, otherwise $K<0$, which makes no physical sense. Hence, the incoherent state in model (\ref{equ:re1}) cannot be linearly stable.
In fact, it is neutrally stable due to the existence of continuous spectrum on the imaginary axis \cite{strogatz1991stability}.
Furthermore, if the coupling strength $K>0$ but sufficiently small, we have proved that the eigenvalue $\lambda$ does not exist (the details are included in the supplementary material).

The analysis above reveals that the linearized operator $\mathcal{T}$ has continuous spectrum $i\omega$ lying on the whole imaginary axis (with real part equals to $0$) for all $K$, and it may also have discrete spectrum (eigenvalues) depending on $K$. When $K$ is small ( $K<K_{c}$ ) the discrete spectrum is empty, but as $K$ increases, discrete eigenvalues emerge with real part $x>0$ for $K>K_{c}$. Imposing the critical condition $x\rightarrow 0^{+}$ for Eq~(\ref{equ:re14}), once again we obtain the critical coupling strength as
\begin{equation}\label{equ:re21}
K_{c}=\dfrac{2}{\pi \textmd{sup}_{j} |y_{j}|g(y_{j})}~,
\end{equation}
where $y_{j}$ are determined by the Eq.~(\ref{equ:re15}) with the limit $x\rightarrow 0^{+}$. Evidently, $\Omega_{c}$ is the imaginary part of the eigenvalues of operator $\mathcal{T}$ at the boundary of stability.  Generally  Eq.~(\ref{equ:re15}) may have more than one root with $x\rightarrow 0^{+}$. $sup_{j}$ means that we choose the $j$th root $y_{j}$ which makes the product $|y|g(y)$ is maximal, so that $K_c$ corresponds to the foremost critical point for the onset of synchronization.

\begin{figure}
  \includegraphics[width=1.05\linewidth,height=0.7\linewidth]{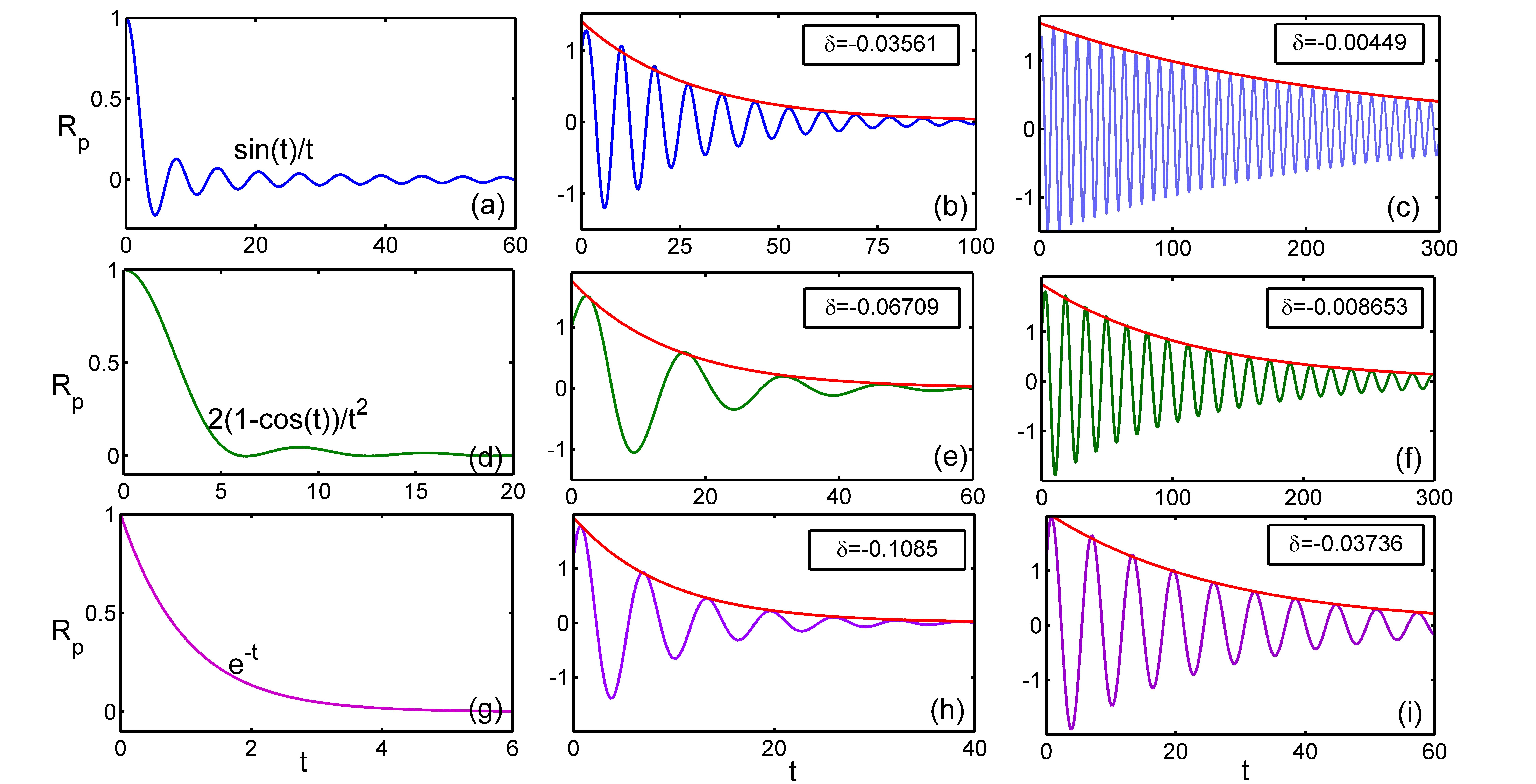}\\
 \caption{Different scenarios of the decay of $R(t)$ with different frequency distributions and coupling strength below the critical threshold ($K<K_{c})$. (a)--(c) Uniform distribution $g(\omega)=\frac{1}{2}, \, \omega\in (-1,1)$. $K=0, 1.6, 1.78$, respectively. (d)--(f) Triangle distribution $g(\omega)=1-|\omega|,\, \omega\in (-1,1)$. $K=0, 2.2, 2.6$, respectively. (g)--(i) Lorentzian distribution $g(\omega)=\frac{1}{\pi}\frac{1}{1+\omega^{2}},\, \omega\in(-\infty,\infty)$. $K=0, 3.7, 3.9$, respectively. The red solid lines denote the fitted curves of the envelopes which all satisfy the exponential form $e^{\delta t}$.
 }\label{fig:1}
\end{figure}

According to the above linear stability analysis, the incoherent state of model (\ref{equ:re1}) is only neutrally stable below the synchronization threshold. However,
interestingly, we find that in this regime the perturbed order parameter $\eta(t)$ actually decays to zero in the long time limit ($t\rightarrow \infty$). This phenomenon was first found in the classical Kuramoto model, and was revealed to be analogous to the famous Landau damping in plasma physics \cite{strogatz1992coupled}. To investigate the Landau damping effect in our model, we rewrite
Eq.~(\ref{equ:re9}) as
\begin{equation}\label{equ:re22}
\dfrac{dZ_{1}}{dt}=i\omega Z_{1}+\dfrac{K}{2}|\omega|\eta(t),
\end{equation}
which could be solved explicitly, i.e.,
\begin{equation}\label{equ:re23}
Z_{1}(t,\omega)=C_{0}e^{i\omega t}+\dfrac{K}{2}|\omega|e^{i\omega t}\int_{0}^{t}e^{-i\omega \tau}\eta(\tau)d\tau.
\end{equation}
Substituting Eq.~(\ref{equ:re23}) into the expression of $\eta(t)$, we obtain the perturbed order parameter as
\begin{equation}\label{equ:re24}
\eta_p (t)=C_{0}\int e^{i\omega t}g(\omega)d\omega+\dfrac{K}{2}\int_{-\infty}^{\infty}|\omega|g(\omega)d\omega\int_{0}^{t}e^{i\omega \tau}\eta_p (t-\tau)d\tau.
\end{equation}
Here $C_{0}$ is a constant related to the initial  value and it is  convenient to set $C_{0}=1$. Eq.~(\ref{equ:re24}) represents a closed form for the dependence of $\eta_p(t)$ on the coupling strength $K$. Unfortunately, it is difficult to get the expression of $\eta_p(t)$ analytically for the present model. However, we still can obtain useful information via direct numerical simulations.
In Fig.~\ref{fig:1}, the numerical solutions of Eq.~(\ref{equ:re24}) are illustrated for different values of $K$ and typical frequency distributions $g(\omega)$. Generally, we observe the decay of $R(t)$, i.e, $|\eta_p(t)|$. Depending on $g(\omega)$ and $K$, the scenarios turn out to be different.
We notice that when the coupling constant is absent, Eq.~(\ref{equ:re24}) can be solved analytically. Specifically,  for example,
when $K=0$,
$R(t)=\sin t/t$ for the uniform distribution [Fig.~\ref{fig:1}(a)],
$R(t)=2(1-\cos t)/t^{2}$ for the triangle distribution [Fig.~\ref{fig:1}(d)], and
$R(t)=e^{-t}$ for the Lorentzian distribution [Fig.~\ref{fig:1}(g)].
In these cases, the decay phenomena strongly depend on the form of $g(\omega)$. However, with the increasing of $K$ the situation changes. It is observed that the order parameter
decays in a way with significant oscillation. Nevertheless, its envelope follows the form of exponential decay, namely, $R(t) \propto e^{\delta t}$ for a short time, where $\delta$ is the decay exponent.
While the general dynamical mechanism of this decay is still an open issue, Ref. \cite{chiba2011center} pointed out that
this exponential decay of order parameter in the neutrally stable regime is closely related to the resonance pole on the left-half complex plane, and the decaying rate $\delta$ is the real part of it.

For the two-cluster synchronous states Eq.~(\ref{equ:re38}), previous analysis has shown that $R=\frac{\sqrt{2}}{2}\sqrt{1+\sqrt{1-4/K^{2}}}$ is linearly stable, and $R=\frac{\sqrt{2}}{2}\sqrt{1-\sqrt{1-4/K^{2}}}$ is linearly unstable~\cite{hu2014exact}.
For the traveling wave state in the range ($0<\alpha<1$), its stability can only be studied through numerical simulations. We have conducted extensive simulations by choosing different initial conditions for phase oscillators. We even specially choose a proper initial condition to make the system artificially locate onto the traveling wave state. It is found that in all these cases the system evolves to the two-cluster synchronous state as long as $K>K_b=2$ (the subscript $b$ denotes the backward transition point). Thus the numerical results give evidence that the traveling wave state predicted by the mean-field theory turns out to be unstable in the current model.

\bigskip
\noindent{\bf The bifurcation analysis.} The above  stability analysis leads to the conclusion that near the finite critical coupling $K_{c}$, the incoherent state becomes unstable with the emergence of a pair of complex conjugated eigenvalues $\lambda_{c}=\pm i\Omega_{c}$; meanwhile the traveling wave solution is unstable. Moreover, due to the absolute coupling, Eq~(\ref{equ:re1}) always has the two-cluster synchronous solution [Eq.~(\ref{equ:re38})] when $K>2$. This is independent of the specific form of $g(\omega)$ as long as $g(\omega)$ is symmetric and centered at $0$ \cite{hu2014exact}. Thus, if $K_{c}>2$, the first-order synchronization transition would take place. However, the mechanism underlying the instability of the incoherent state is still unclear, for example, the bifurcation type and the local stability of the traveling wave solution near $K_{c}$. These information is crucial for us to get a global picture of the synchronization transition in the dynamical system.
Generally, the dynamic behavior near the critical point can be investigated through the local bifurcation theory. For this purpose, we refer to the  framework of nonlinear analysis developed in Ref. \cite{crawford1994amplitude} to reveal the local bifurcation type for the incoherent state of model (\ref{equ:re1}).

The main idea of the theory~\cite{crawford1994amplitude} is that when the perturbed equation of the incoherent state satisfies $O(2)$ symmetry, the center manifold reduction could be applied to obtain the amplitude equations for both steady state and limit cycle. Moreover, in order to avoid dealing with the continuous spectrum, the Gaussian white noise is added and eventually the noise magnitude is extended to zero for all calculations. Following this treatment, now the evolution of the density function $\rho(\theta,t,\omega )$ obeys the Fokker-Planck equation
\begin{equation}\label{equ:re41}
\dfrac{\partial\rho}{\partial t}+\dfrac{\partial}{\partial \theta}(\rho\cdot\upsilon)=D\dfrac{\partial^{2}\rho}{\partial \theta^{2}},
\end{equation}
where $D$ is the strength of noise. Similarly, imposing the small perturbation to the incoherent state, i.e., $\rho=\frac{1}{2\pi}+\mu(\theta,\omega,t)$, we obtain the following perturbed equation
\begin{equation}\label{equ:re42}
\dfrac{\partial\mu}{\partial t}=\mathscr{L}\cdot \mu+\mathscr{N}(\mu).
\end{equation}
Here, $\mathscr{L}\cdot \mu$ is lineared equation for $\mu$, i.e.,
\begin{equation}\label{equ:re43}
\mathscr{L}\cdot\mu=D\dfrac{\partial^{2}}{\partial \theta^{2}}\mu-\omega\dfrac{\partial\mu}{\partial \theta}+\dfrac{K|\omega|}{2}\cdot\dfrac{1}{2\pi}\left[e^{-i\theta}Z+e^{i\theta}Z^{*}\right],
\end{equation}
and $\mathscr{N}(\mu)$ is the nonlinear term, i.e.,
\begin{equation}\label{equ:re44}
\mathscr{N}(\mu)=\dfrac{K|\omega|}{2}\left[e^{i\theta}(\mu-i\dfrac{\partial\mu}{\partial \theta})Z^{*}+e^{-i\theta}(\mu+i\dfrac{\partial\mu}{\partial \theta})Z\right],
\end{equation}
where $Z=\int_{0}^{2\pi}e^{i\theta'}d\theta'
\int_{-\infty}^{\infty}\mu(\theta',\omega',t)g(\omega')d\omega'$. Then we can derive the normal form of amplitude equation for both the steady state and the Hopf bifurcation in the frequency-weighted model based on the center manifold assumption. Since the complete calculation is tedious, we put the details into the supplementary material for interested readers. In the following we only report the main results.

It is found that for the frequency-weighted model, the system undergoes Hopf bifurcation near the critical coupling $K_{c}$. As a result, a traveling wave solution and a standing wave solution emerge above $K_{c}$.
The standard standing wave solution consists of
two counter-rotating clusters of phase-locked oscillators.
Thus  the order parameter $\eta(t)$ plots
a limit cycle on the complex plane.
Previously,
such state has been found  in the classical Kuramoto model with symmetric bimodal frequency distribution \cite{okuda1991mutual,bonilla1992nonlinear,bonilla1998time}.
It should be pointed out that the mean-field theory fails to
predict such state due to the fact that
neither the distribution function nor $R(t)$ are stationary in any rotating frame for such a state.

\begin{figure}
  \includegraphics[width=0.9\linewidth,height=0.6\linewidth]{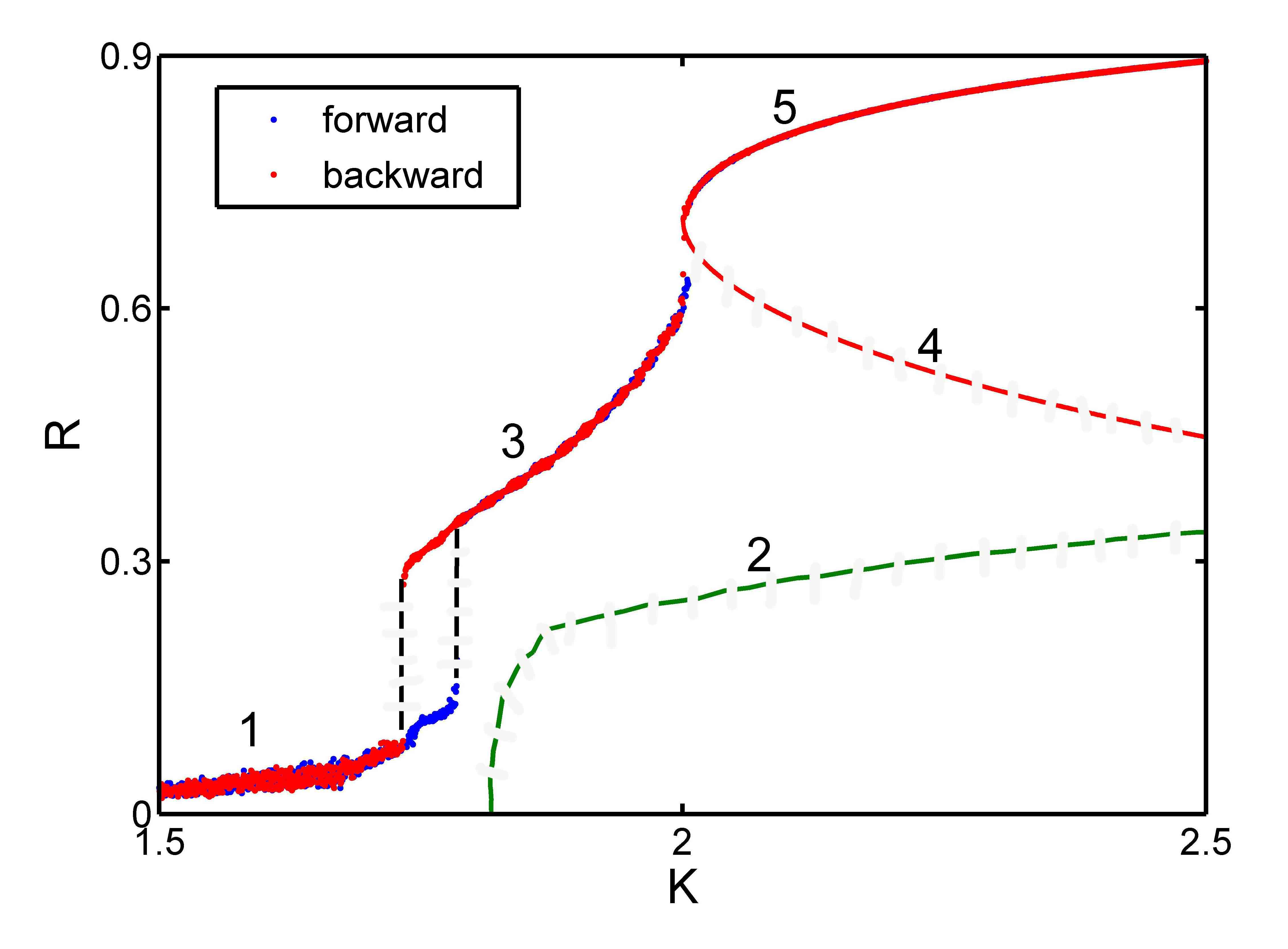}\\
  \caption{ Characterizing various coherent states in the phase diagram. $R$ [long time average of $R(t)$] vs. $K$ for the uniform frequency distribution $g(\omega)=\frac{1}{2}, \, \omega\in (-1,1)$. Branches $1--5$ are the incoherent state, the (unstable) traveling wave state predicted by the mean-field theory, the standing wave state, the unstable and the stable two-cluster synchronous states, respectively . The blue and red lines denote the forward and the backward transitions, respectively. In both directions, $K$ is changed adiabatically in simulations. There is a hysteresis region of the standing wave solution within $K=1.725-1.8$. In the simulations
oscillators number $N= 50, 000$, and a fourth-order Runge-Kutta integration method with time step 0.01 is used. }\label{fig:2}
\end{figure}

As an example to illustrate our results, we focus on the case of uniform distribution $g(\omega)=1/2$. The exact expression for critical coupling strength is $K_{c}=4\sqrt{2}/\pi<2$.
The nonlinear analysis shows that the bifurcation for the traveling wave solution is supercritical and unstable (which is consistent with the mean-field theory). In addition, the standing wave solution is subcritical. This implies that a hysteresis would occur by taking the high order terms of the amplitude equation into account. Numerical evidence suggests that above $K_{c}$ the incoherent state loses its stability. Meanwhile,   non-stationary $R(t)$ emerges with a hysteresis near $K_{c}$ [branch 3 in Fig.~\ref{fig:2}].
As the  coupling strength increases, and it eventually vanishes at $K=2$ via a discontinuous transition (with very small hysteresis loop) to the two-cluster synchronous state. We have also conducted calculations for other typical frequency distributions, such as the triangle, the Lorentzian, and the parabolic. The results show that the bifurcations for the standing wave solution are all subcritical and the traveling wave solution are all unstable locally. Moreover,  the direction of bifurcation for the traveling wave state supports the numerical solution of the mean-field equation.
It should be pointed out that the stable branch of subcritical bifurcation for both states are not observed numerically.
One possible reason for this is that their basins of attraction might be so small in such a high-dimensional phase space that most of the initial conditions eventually lead to  the stable two-cluster synchronous state as long as $K>2$.

\bigskip
\noindent{\large\bf Discussion}
\\ \noindent
To summarize, we investigated the  synchronization transition in the frequency-weighted Kuramoto model with all-to-all couplings. Theoretically,  mean-field analysis, linear stability analysis, and  bifurcation analysis have been carried out to obtain insights.
Together with the numerical simulations, our study presented the following main results.
First,
we predicted the possible steady states in this model, including the  incoherent state, the two-cluster synchronous state, the traveling wave state, and the standing wave state.
Second,
The  critical coupling strength for synchronization transition has been obtained analytically.
Third,
We proved that in this model
the incoherent state is only neutrally stable below the synchronization threshold. However, in this regime, the perturbed order parameter decays exponentially to zero for short time.
Finally, the amplitude equations near the bifurcation point have been derived based on the center-manifold reduction, which
predicted that the non-stationary standing wave state could also exist in this model.
This work provided a complete framework to deal with the frequency-weighted Kuramoto model, and the obtained results will enhance our understandings of the first-order synchronization transition in networks.

\section*{Acknowledgements}
This work is partially supported by the NSFC grants Nos. 11075016,  11475022, 11135001, and the Scientific Research Funds of Huaqiao University.

\section*{Author contributions}
C.X., Y.T.S., J.G., T.Q, S.G.G and Z.G.Z. designed the research; C.X., Y.T.S. and T.Q performed numerical simulations and theoretical analysis; C.X., S.G.G and Z.G.Z. wrote the paper. All authors reviewed and approved the manuscript.

\section*{Additional information}
{\bf Competing financial interests:} The authors declare no competing financial interests.\\

Correspondence and requests for materials should be addressed to Z.G.Z. (zgzheng@bnu.edu.cn), or S.G.G. (guanshuguang@hotmail.com).


\clearpage

\clearpage

\begin{table}
\small
  \centering
  \caption{Summary of the frequency distributions, the balance equations,  the critical mean-field frequencies $\Omega_{c}$, and the critical coupling strength $K_{c}$.
  From top to bottom: the uniform distribution,
  the triangle distribution,
  the parabolic distribution,
  and the Lorentzian distribution.}\label{table:1}
  \begin{tabular}{|c|c|c|c|}
  \hline
     frequency distribution & balance equation & \textit{$\Omega_{c}$} & \textit{$K_{c}$} \\
     \cmidrule(r){1-1}     \cmidrule(lr){2-2}   \cmidrule(lr){3-3} \cmidrule(lr){4-4}
     \hline
     $g(\omega)=\frac{1}{2a}\Theta(a-|\omega|)$ & $\frac{\Omega_{c}}{2a}\textmd{ln}\frac{a^{2}-\Omega_{c}^{2}}{\Omega_{c}^{2}}=0$ & $0,\; \pm\frac{a}{\sqrt{2}}$ & $\frac{4\sqrt{2}}{\pi}$ \\
     \cmidrule(r){1-1}     \cmidrule(lr){2-2}   \cmidrule(lr){3-3} \cmidrule(lr){4-4}
     \hline
     $g(\omega)=\frac{a-|\omega|}{a^{2}}\Theta(a-|\omega|)$ & $\frac{\Omega_{c}}{a^{2}}[2a+(\Omega_{c}-a)\textmd{ln}\frac{a-\Omega_{c}}{\Omega_{c}}+(a+\Omega_{c})\textmd{ln}\frac{\Omega_{c}}{a+\Omega_{c}}]=0$ & $0,\; \pm0.40$ & $2.65$ \\
     \cmidrule(r){1-1}     \cmidrule(lr){2-2}   \cmidrule(lr){3-3} \cmidrule(lr){4-4}
     \hline
     $g(\omega)=\frac{3}{4a^{3}}(a^{2}-\omega^{2})\Theta(a-|\omega|)$ & $\frac{3\Omega_{c}}{4a^{3}}[(a^{2}-\Omega_{c}^{2})\textmd{ln}\dfrac{a^{2}-\Omega_{c}^{2}}{\Omega_{c}^{2}}-a^{2}]=0$ & $0,\;\pm 0.47$ & $2.34$ \\
     \cmidrule(r){1-1}     \cmidrule(lr){2-2}   \cmidrule(lr){3-3} \cmidrule(lr){4-4}
     \hline
     $g(\omega)=\frac{1}{\pi}\frac{\gamma}{\omega^{2}+\gamma^{2}}$ & $\frac{2\gamma}{\pi}
     \Omega_{c}\textmd{ln}\frac{\gamma}{\Omega_{c}}/(\gamma^{2}+\Omega_{c}^{2})=0$ & $0,\;\pm\gamma$ & 4 \\
     \hline
   \end{tabular}
\end{table}

\end{document}